\documentclass{article}

\PassOptionsToPackage{square, numbers, compress}{natbib}
\usepackage[preprint]{neurips_2019}
\usepackage[utf8]{inputenc} 
\usepackage[T1]{fontenc}    
\usepackage{microtype}
\usepackage{graphicx}
\usepackage{subfigure}
\usepackage{booktabs, tabularx}
\usepackage{hyperref}
\usepackage{url}  
\usepackage{amsthm}
\usepackage{bbm}
\usepackage{amssymb}
\usepackage{amsmath,bm}
\usepackage{amsfonts}
\usepackage{dsfont}
\usepackage{nicefrac}
\usepackage{algorithm}
\usepackage{algorithmic}
\usepackage{xcolor}

\newcommand{\sA}{{\mathcal{A}}}
\newcommand{\sB}{{\mathcal{B}}}
\newcommand{\sC}{{\mathcal{C}}}

\newcommand{\sE}{{\mathcal{E}}}

\newcommand{\sO}{{\mathcal{O}}}

\newcommand{\sS}{{\mathcal{S}}}

\newcommand{\sX}{{\mathcal{X}}}

\newcommand{\bK}{{k}}

\title{Re-ranking Based Diversification: A Unifying View}
\author{
    Shameem A Puthiya Parambath\\
    QCRI, Doha, Qatar\\
    \texttt{spparambath@hbku.edu.qa}
}

\begin{document}

\maketitle

\begin{abstract}
We analyze different re-ranking algorithms for diversification and show that majority of them are based on maximizing submodular/modular functions from the class of parameterized concave/linear over modular functions.
We study the optimality of such algorithms in terms of the `total curvature'.
We also show that by adjusting the hyperparameter of the concave/linear composition to trade-off relevance and diversity, if any, one is in fact tuning the `total curvature' of the function for relevance-diversity trade-off.
\end{abstract}

\section{Introduction}
Standard personalized recommendation algorithms consider only relevance of the items to a user when making predictions.
This strategy implicitly assumes that relevance of the items are independent of each other.
In practice this draws in similar and redundant items to be recommended.
Moreover, popularity bias in recommender systems makes the situation worse by excluding unpopular but relevant items from recommendations.
In personalized recommendation, diversification is a means to recommend novel, serendipitous items that result in higher user satisfaction \citep{ziegler2005improving,ge2010beyond,vargas2014coverage,puthiya2016coverage} whereas in group recommendation, diversification is a way to generate consensus recommendations by finding items relevant to a group of diverse users \citep{puthiya2018saga}.

A common strategy for diversification is re-ranking: given a list of recommendations, re-ranking algorithms reorder the list to account for diversity \citep{hurley2013personalised}.
Recently many re-ranking based diversification algorithms have been proposed which exploit different aspects of the recommendations like personal popularity tendency \citep{oh2011novel}, genre coverage \citep{vargas2014coverage}, interest coverage \citep{puthiya2016coverage} etc.
It is already shown that the `diminishing return' property of submodular functions enable the system to trade-off between relevance and diversity \citep{chapelle2011intent}. 
A function $F$ defined on the subsets of a ground set $\sE$ is called submodular if for all subsets $\sA,\sB \subseteq \sE$, 
\begin{equation}
F(\sA) + F(\sB) \geq F(\sA\cup \sB) + F(\sA\cap \sB)
\label{eq:sub}
\end{equation}
$F$ is modular if strict equality holds in \eqref{eq:sub}.
Though not very apparent, careful analysis reveals that objective functions corresponding to these algorithms are not only submodular/modular but take a simple form: the composition of a concave/linear function with a modular function.

The purpose of this paper is not to propose a new diversification algorithm; we aim to uncover the exact functional forms of submodular or modular functions used in re-ranking based diversification algorithms and see how this affects the relevance-diversity trade off.
We analyze objective functions for many re-ranking algorithms and show that they belong to the the class of concave/linear over modular functions.
We also study the relevance-diversity trade-off of re-ranking algorithms and establish that relevance-diversity trade-off of these algorithms is tied to the `total curvature' of the objective function.
By changing the hyperparameters associated with the objective function, one is effectively tuning the `total curvature' of the objective.

The remainder of this paper is structured as follows.
We give a brief overview of diversification algorithms in Section~\ref{sec:rel_work}.
In Section~\ref{sec:div_alg}, we show that popular diversification algorithms are based on the class of parameterized concave/linear composition of modular functions.
We also discuss the `total curvature' and the worst case lower bound for submodular maximization in terms of the `total curvature'.
We discuss our experimental results in Section~\ref{sec:exp} before concluding the paper in Section~\ref{sec:conc}.

\section{Related Work}
\label{sec:rel_work}
Recommendation diversification problem has been studied extensively in the past \citep{onuma2009tangent,oh2011novel,vargas2014coverage,puthiya2016coverage,wasilewski2018,cheng2017learning,hurley2013personalised}.
Broadly speaking, diversification algorithms come under two schemes: \emph{(i)} re-ranking based algorithms \emph{(ii)} multi-objective optimization based algorithms.
Re-ranking based diversification algorithms or simply re-ranking algorithms are the major point of discussion here.
Given a recommendation list produced by any personalized recommendation algorithms, re-ranking algorithms re-order the items such that relevant and diverse items appear in the top-$\bK$ rankings.
Re-ranking algorithms are easy to implement and popular because they can be readily plugged in to existing personalized recommender systems as a post-processing step.

In the seminal work by \citet{carbonell1998use}, the authors introduced MMR, a method to re-rank web search results.
A number of papers followed, especially in specific application settings like web search \citep{santos2010exploiting} and document summarization \citep{lin2011class}.
Several re-ranking algorithms have been proposed in recommender systems \citep{onuma2009tangent,oh2011novel,su2013set,vargas2014coverage,Wu2016RMC,puthiya2016coverage,wasilewski2018}.
A notable characteristic of these algorithms is that the objective function can be represented as a modular/submodular maximization problem with matroid constraints and a solution can be obtained using a simple greedy heuristic.
Moreover, the celebrated result due to \citet{nemhauser1978analysis} states that steepest ascent greedy algorithm guarantees a constant approximation worst case lower bound i.e. $F(\sS^{\ast}) \geq (1-\frac{1}{e})F(\sS^{opt})$ where $e$ is the base of the natural logarithm, $\sS^{\ast}$ is the greedy solution and $\sS^{opt}$ is the unknown optimal solution.
We analyze these algorithms in detail in the following sections. 

In multi-objective optimization based algorithms, a model optimized for an objective function comprising of both relevance and diversity is used to predict the recommendations.
\citep{hurley2013personalised} proposed an objective which combines the standard latent factor model with intra-list distance, a measure of diversity.
Similarly, \citet{cheng2017learning} proposed diversified collaborative filtering to learn a prediction model for diverse personalized ranking. 
In addition, there has been some work on unifying the performance metrics for diverse recommendations.
\citet{ziegler2005improving} proposed intra-list distance as a measure of diversity of the recommendation.
In \citep{ge2010beyond}, authors argued that accuracy alone does not capture the "fitness of use" of the recommendation and that further metrics are required to capture non-trivial serendipitous recommendations which increases user satisfaction.
A serendipity measure based on average popularity of the items is proposed in \citep{ziegler2005improving}.

\section{Diversification Algorithms}
\begin{table}
    \begin{tabularx}{\columnwidth}{@{\hskip 1in}l@{\hskip 1in}l@{\hskip 1in}l@{\hskip 1in}}
    \toprule
            Algorithm &   $g(x)$  \\
            \midrule
            \citet{carbonell1998use}     & $x$  \\
            \citet{onuma2009tangent}     & $\frac{1}{x}, \, x>0$  \\
            \citet{oh2011novel}          & $\log(x), \, x>0$ \\
            \citet{su2013set}            & $\frac{1}{\lambda x}, \, w,x>0$ \\
            \citet{vargas2014coverage}   & $x^{\lambda},\, \lambda \in [0,1]$ \\
            \citet{puthiya2016coverage}  & $x^{\lambda},\, \lambda \in [0,1]$ \\
            \citet{Wu2016RMC}            & $\frac{x}{1+x}$ \\
            \citet{wasilewski2018}       & $\lambda x, \, \lambda >0$ \\
            \bottomrule
    \end{tabularx}
    \caption{Functional form of $g(x)$ in re-ranking algorithms}
    \label{tab:1}
    \label{tab:res1}
\end{table}
\label{sec:div_alg}
Before we start analyzing different diversification algorithms, we introduce our notation and some basic definitions.
Our discussion lies primarily within the collaborative filtering domain.
We denote the set of items using $\sX$, set of observed (rated) items using $\sO$ and $\sE$ denotes the set of unobserved items.
$\sS$ denotes a subset of $\sE$ such that $|\sS| \leq \bK$.
$rel(i)$ indicates the relevance of an item $i$ and $rel(\sS)$ indicates the relevance of the set $\sS$, often defined as the sum of the monotone transformation of relevance of individual items.

Below we discuss some seminal work on re-ranking and show that these algorithms are based on maximizing a submodular/modular objective function from the class of parameterized concave/linear composition of modular functions.
Maximal Marginal Relevance (MMR) \citep{carbonell1998use} is one of the earliest re-ranking algorithms with objective function
\begin{equation*}
F(\sS) = \sum_{i \in \sS} \Big(\lambda*sim_1(u,i) - (1-\lambda) \max \limits_{j \in \sS\setminus \{i\}} sim_2(i,j)\Big)
\end{equation*}
where $sim_1(u,i)$ is a modular function representing the relevance of a user $u$ to the item $i$ and similarly $sim_2(i,j)$ is also a modular function representing the similarity between items $i$ and $j$.
$\lambda$ in the above formulation is the hyperparameter that is tuned for relevance-diversity trade-off.
Hence $F(\sS)$ is in fact the difference of two modular functions i.e a linear composition of modular functions and is shown to be submodular \citep{lin2011class}.
Many re-ranking algorithms inspired from MMR were proposed in specific application settings like web search diversification \citep{santos2010exploiting}.
Those algorithms follow the same functional form as in the case of MMR.
In fact, many re-ranking algorithms like MMR take a simple form as given below
\begin{equation}
F(\sS) = f(S) + \lambda g(h(S))
    \label{eq:gen}
\end{equation}
where $f$ is a modular function representing the relevance of recommendation set $\sS$, $g(h)$ is the diversity term which is the composition of a linear or concave function, $g$, and a modular function $h$ and $\lambda$ is the hyperparameter.
Often $h$ is defined as a function of relevance itself, but capturing diversity aspects like coverage, popularity, serendipity etc.

TANGENT algorithm \citep{onuma2009tangent} proposed an objective function similar to the MMR objective.
The objective function can be written as $\displaystyle{F(\sS) = rel(S) + \frac{1}{rel(S)}}$.
Here $f$ is simply the sum of relevance of items and $g$ is the concave reciprocal function.
The objective function proposed by \citet{oh2011novel} takes the form $\displaystyle {F(\sS) = rel(\sS) +\log{\frac{h(\sS)}{h(\sO)}}}$.
Here $h(\sS)$ is defined as the \emph{personal popularity tendency} of the set $\sS$ with respect to a user.
Here diversity term is defined as the composition of the concave function $\log$ with the modular function $\frac{h(\sS)}{h(\sO)}$.
BinomDiv\citep{vargas2014coverage} also re-ranks the recommendations by defining a diversity term for genre coverage.
Like in the previous case, $f$ corresponds to a modular function indicating the relevance of an item and diversity term is defined as the product of coverage and non-redundancy.
Both coverage and non-redundancy terms are defined in terms of the genre coverage and take the form $x^{\lambda}$ with $\lambda \in [0,1]$.
The objective function proposed in \citep{Wu2016RMC} also consists of two terms, the first modular term captures the relevance with respect to neighbouring users and the second term, the neighbour coverage function stands as a surrogate for diversity.
It is defined as the concave composition $g(x) = \frac{x}{1+x}$ of a positive modular function.

It is also very common that only the diversity aspect is considered for re-ranking i.e. $f(\sS) = 0$ in \eqref{eq:gen}.
\citet{su2013set} proposed a re-ranking algorithm by considering only the sub-categories associated with items.
The objective function considers only diversity aspect and takes the form $F(\sS) = \sum_{i\in \sC(\sS)} \sum_{j \in \sC(\sS_{-i})} \frac{1}{|\sC(\sS)||\sC(\sS_{-i})|}$ where $\sC(\sS)$ represents the categories associated with the set of items $\sS$ and $\sS_{-i} = \sS \setminus \{i\}$.
This is equivalent to the sum of the concave function $g(x) = \nicefrac{1}{x}$. 
\citet{puthiya2016coverage} also proposed a re-ranking algorithm considering only the diversity term.
The diversity term takes the form $x^\lambda,\, \lambda \in [0,1]$ and is defined in-terms of the interest coverage.
In \citep{wasilewski2018}, the authors proposed an intent-aware diversification algorithm.
\citet{chapelle2011intent} showed that intent-aware objective functions are either submodular or modular functions.
The objective function in \citep{wasilewski2018} is modular since it is based on re-weighting item based collaborative filtering scores.
Table:\ref{tab:1} summarizes the essence of the above discussion regarding re-ranking algorithms.

\subsection{Diversity-Relevance Trade-off}
We analyze the relevance-diversity trade-off in re-ranking algorithms to get a deeper theoretical understanding.
In majority of the re-ranking algorithms, the trade-off in relevance and diversity is obtained by tuning the associated hyperparameter.
For example, in case of MMR \citep{carbonell1998use} based algorithms $\lambda$ is the hyperparameter controlling the trade-off.
Similarly in \citep{vargas2014coverage,puthiya2016coverage}, $\lambda$ parameter controls the relevance-diversity trade-off.
In further discussions, we consider the re-ranking algorithm proposed in \citep{puthiya2016coverage} as a use case.

The total curvature of a non-decreasing submodular set function with respect to a set $\sS$ is defined as
\begin{equation}
    \alpha = \max_{j\in \sS} \frac{F(\sS\setminus \{j\}) + F(j) - F(\sS)}{F(j)}
\label{eq:tot_curv}
\end{equation}
Intuitively, total curvature measures how far $F$ is from being modular and \eqref{eq:tot_curv} represents the distance of a monotone submodular function to modularity.
Total curvature can take values between 0 and 1, and it is zero in case of modular functions and one in case of matroid rank function.
\citet{conforti1984submodular} extended the result in \citep{nemhauser1978analysis} and gave a tighter lower bound for the submodular maximization problem  in terms of total curvature.
According to \citep{conforti1984submodular}, $F(\sS^{\ast}) \geq \frac{1}{\alpha}(1-e^{-\alpha})F(\sS^{opt})$.

The curvature can be computed very easily for any submodular function, assuming a value oracle model.
When the curvature is 1, this gives the standard approximation bound given in \citep{nemhauser1978analysis} and for any other values it strengthens the approximation bound in \citep{nemhauser1978analysis}.
For example, if the curvature value $\alpha = 0.1$, we are guaranteed that $F(S^{\ast}) \geq 0.95F(S^{opt})$ compared to the standard result of $F(S^{opt}) \geq 0.63F(S^{\ast})$.
It should be noted that curvature does not rely on specific functional form of $F$ but only on the marginal gains.

We argue that by changing the hyperparameter, one effectively changes the total curvature of the re-ranking objective.
One can adjust the parameter such that the re-ranking objective is simply a modular function.
For example, by setting $\lambda = 1$ in \citep{carbonell1998use} or setting $\lambda = 1$ in \citep{vargas2014coverage,puthiya2016coverage},
total curvature of the resulting modular function can be made 0, and the recommendation list will be least diverse containing mostly relevant items.
Similarly, by adjusting the hyperparameter such that total curvature is 1, one can obtain the most diverse set of recommendations.

\section{Experiments}
\label{sec:exp}
In this section, we experimentally validate our claims on a movie recommendation task. 
We followed the same experimental protocol as given in \citep{puthiya2016coverage}.
We used the benchmark MovieLens 1M dataset and carried out holdout validation by splitting the data into training and test set such that 5\% of the original data goes into testing and the remaining goes into training.
The reported results are the average over the five splits.

We used regularized weighted non-negative matrix factorization to extract the user and item features \cite{steck2013evaluation}.
The extracted user/item features are used to estimate the user-user/item-item similarity matrix.
We used the diversity objective proposed in \citep{puthiya2016coverage}. 

\subsection{Evaluation Metrics}
The performance of diverse ranking tasks are evaluated using three metrics: a serendipity metric and a dissimilarity metric to measure the distinctiveness of the recommendations and Discounted Cumulative Gain (DCG) as the ranking metric.
DCG is a binary ranking metric and it is calculated by discretizing the observed rating values such that rating values of 4 and 5 are deemed as relevant and as irrelevant otherwise.
We define serendipity score (SS) as the inverse of the average popularity of the recommended items which are not rated by the user.
    SS is a good indicator of the 'non-obvious' recommendations \citep{ziegler2005improving}.
We used feature distance (FD) as a measure of dissimilarity between the recommended items, and is defined as the average euclidean distance between the item feature vectors in the recommended set and it is has been used as a diversity objective in the past \citep{cheng2017learning}.

\subsection{Results \& Discussion}
\begin{figure}
    \begin{center}
        \begin{tabular}{c}
            \includegraphics[width=0.70\columnwidth]{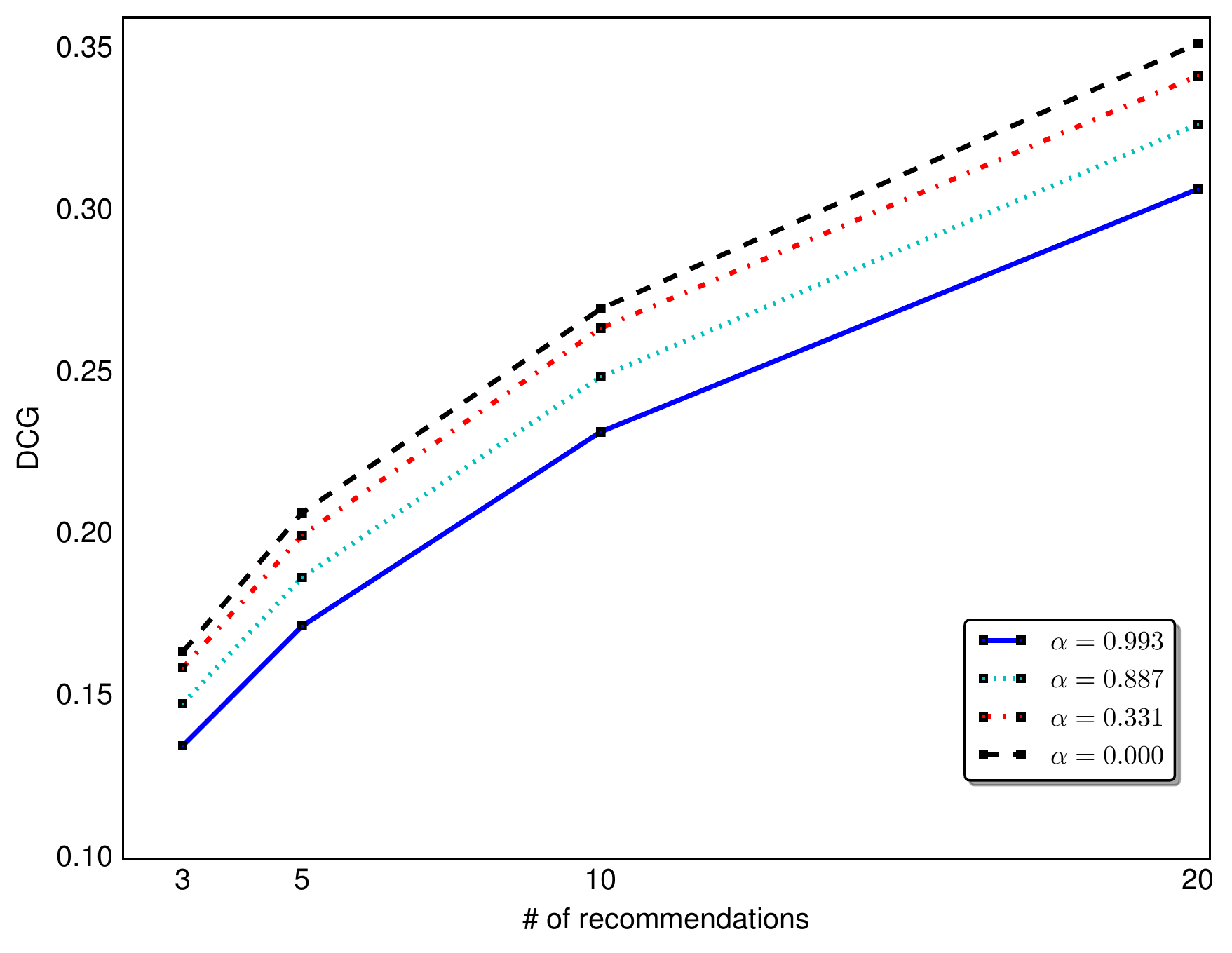} \\ 
            \includegraphics[width=0.70\columnwidth]{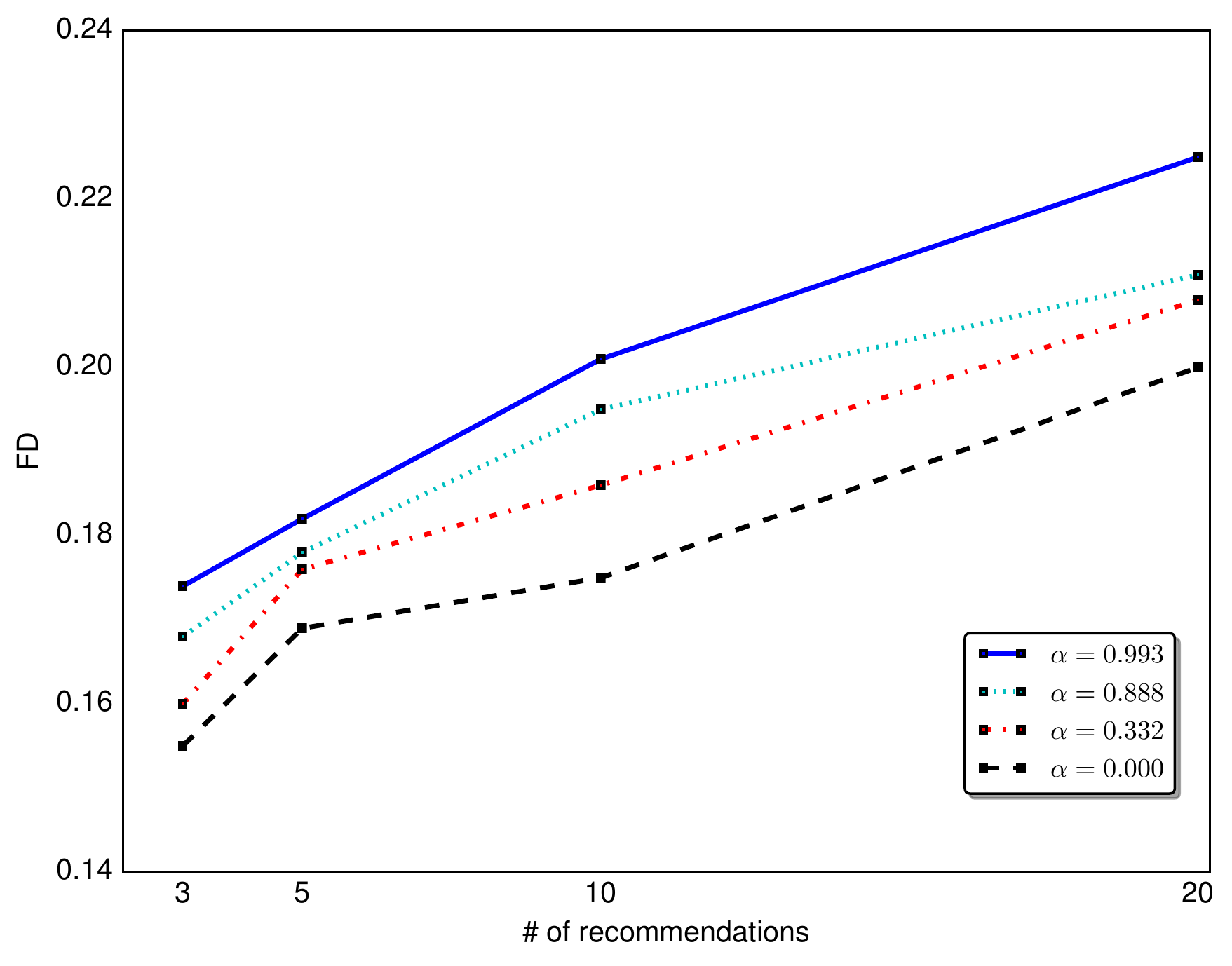} \\
            \includegraphics[width=0.70\columnwidth]{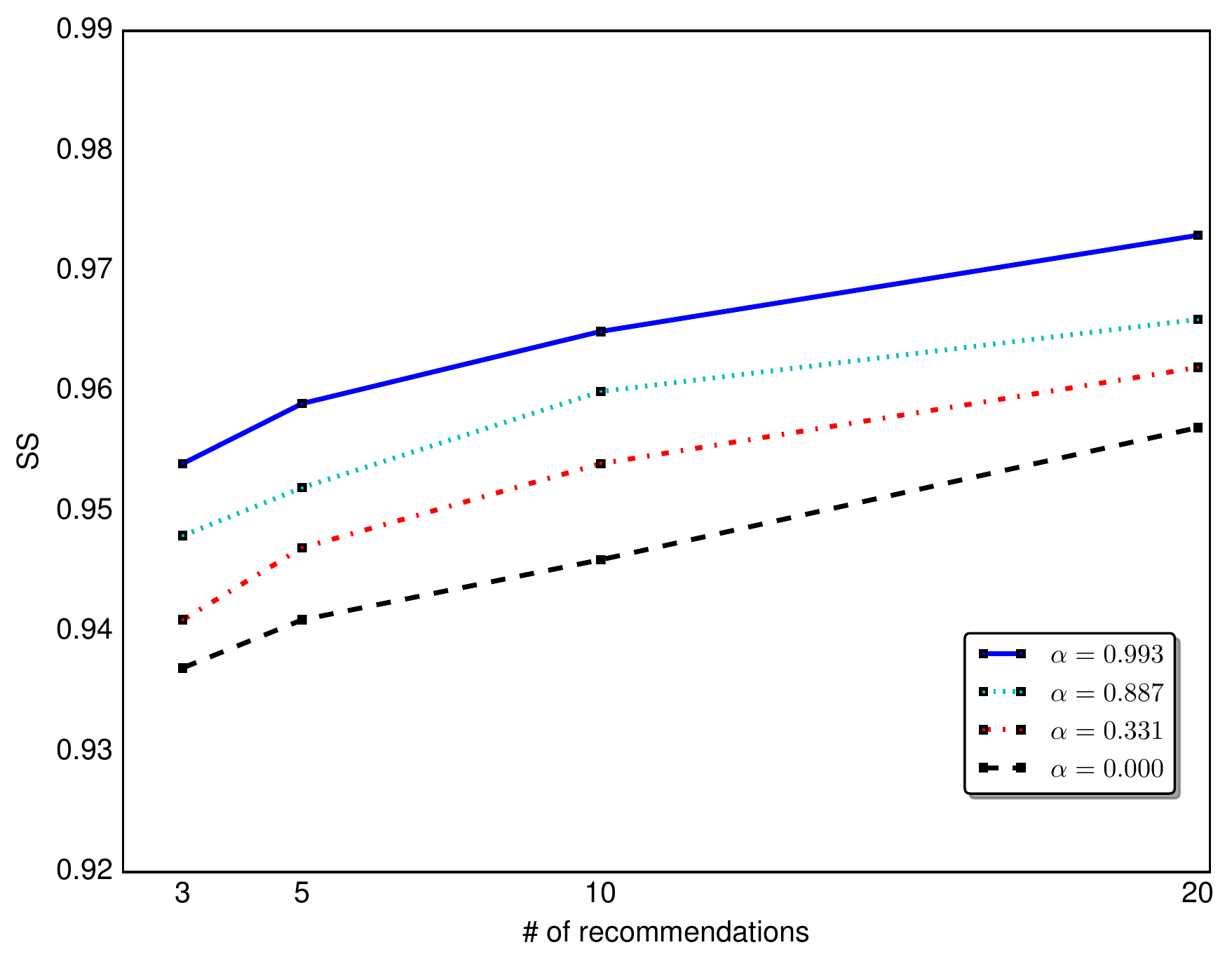} \\
        \end{tabular}
    \end{center}
    \caption{Relevance-Diversity trade-off as a function of recommendation size for different values of total curvature ($\alpha$)}
    \label{fig:results1}
\end{figure}

The results of our experiments are plotted in Fig~\ref{fig:results1}.
For any value of $\alpha$ less than 1, greedy algorithm returns a mix of diverse and relevant items with the lower bound guarantee provided by the corresponding $\alpha$ value.

As the total curvature ($\alpha$) value increases from 0 to 1, DCG value decreases.
For $\alpha = 0$, the re-ranking objective becomes a pure modular function and the greedy algorithm returns recommendations containing the most relevant items as indicated by the highest DCG value.
As $\alpha$ value approaches 1, the objective becomes `easy' to `difficult', and many diverse items are added to the recommendation list.
A modular objective is `easy' as the solution obtained using greedy heuristic is optimal whereas submodular objective is `difficult' as the greedy solution is sub-optimal.
Consequently, SS and FD values increases as the value of $\alpha$ increases from 0 to 1, sacrificing DCG values.

As argued by \citet{ziegler2005improving}, diversification approach bears particularly high impact in recommendation with sequential consumption schemes like online video or music streaming.
There should be a right mix of items which are diverse yet relevant to the user.
By adjusting the hyperparameter of the re-ranking algorithms such that total curvature value is closer to 0.5, one can hopefully get the right mix of items for consumption.
\section{Conclusion}
\label{sec:conc}
We analyzed different diversification algorithms and have shown that majority of these algorithms are based on maximizing a submodular/modular objective function from the class of parameterized concave/linear over modular functions.
We also showed that by varying the parameter of the concave/linear composition, one effectively tunes the `total curvature' of the objective for relevance-diversity trade-off.
\bibliographystyle{plainnat}
\bibliography{recsys}

\begin{thebibliography}{19}
\providecommand{\natexlab}[1]{#1}
\providecommand{\url}[1]{\texttt{#1}}
\expandafter\ifx\csname urlstyle\endcsname\relax
  \providecommand{\doi}[1]{doi: #1}\else
  \providecommand{\doi}{doi: \begingroup \urlstyle{rm}\Url}\fi

\bibitem[Carbonell and Goldstein(1998)]{carbonell1998use}
Jaime Carbonell and Jade Goldstein.
\newblock The use of {MMR}, diversity-based reranking for reordering documents
  and producing summaries.
\newblock In \emph{SIGIR}, pages 335--336. ACM, 1998.

\bibitem[Chapelle et~al.(2011)Chapelle, Ji, Liao, Velipasaoglu, Lai, and
  Wu]{chapelle2011intent}
Olivier Chapelle, Shihao Ji, Ciya Liao, Emre Velipasaoglu, Larry Lai, and
  Su-Lin Wu.
\newblock Intent-based diversification of web search results: metrics and
  algorithms.
\newblock \emph{Information Retrieval}, 14\penalty0 (6):\penalty0 572--592,
  2011.

\bibitem[Cheng et~al.(2017)Cheng, Wang, Ma, Sun, and Xiong]{cheng2017learning}
Peizhe Cheng, Shuaiqiang Wang, Jun Ma, Jiankai Sun, and Hui Xiong.
\newblock Learning to recommend accurate and diverse items.
\newblock In \emph{WWW}, pages 183--192. International World Wide Web
  Conferences Steering Committee, 2017.

\bibitem[Conforti and Cornu{\'e}jols(1984)]{conforti1984submodular}
Michele Conforti and G{\'e}rard Cornu{\'e}jols.
\newblock Submodular set functions, matroids and the greedy algorithm: tight
  worst-case bounds and some generalizations of the rado-edmonds theorem.
\newblock \emph{Discrete applied mathematics}, 7\penalty0 (3):\penalty0
  251--274, 1984.

\bibitem[Ge et~al.(2010)Ge, Delgado-Battenfeld, and Jannach]{ge2010beyond}
Mouzhi Ge, Carla Delgado-Battenfeld, and Dietmar Jannach.
\newblock Beyond accuracy: evaluating recommender systems by coverage and
  serendipity.
\newblock In \emph{RecSys}, pages 257--260. ACM, 2010.

\bibitem[Hurley(2013)]{hurley2013personalised}
Neil~J Hurley.
\newblock Personalised ranking with diversity.
\newblock In \emph{Proceedings of the 7th ACM conference on Recommender
  systems}, pages 379--382. ACM, 2013.

\bibitem[Lin and Bilmes(2011)]{lin2011class}
Hui Lin and Jeff Bilmes.
\newblock A class of submodular functions for document summarization.
\newblock In \emph{ACL}, pages 510--520, 2011.

\bibitem[Nemhauser et~al.(1978)Nemhauser, Wolsey, and
  Fisher]{nemhauser1978analysis}
George~L Nemhauser, Laurence~A Wolsey, and Marshall~L Fisher.
\newblock An analysis of approximations for maximizing submodular set
  functions.
\newblock \emph{Mathematical Programming}, 14\penalty0 (1), 1978.

\bibitem[Oh et~al.(2011)Oh, Park, Yu, Song, and Park]{oh2011novel}
Jinoh Oh, Sun Park, Hwanjo Yu, Min Song, and Seung-T Park.
\newblock Novel recommendation based on personal popularity tendency.
\newblock In \emph{ICDM}, pages 507--516. IEEE, 2011.

\bibitem[Onuma et~al.(2009)Onuma, Tong, and Faloutsos]{onuma2009tangent}
Kensuke Onuma, Hanghang Tong, and Christos Faloutsos.
\newblock Tangent: a novel,'surprise me', recommendation algorithm.
\newblock In \emph{Proceedings of the 15th ACM SIGKDD international conference
  on Knowledge discovery and data mining}, pages 657--666. ACM, 2009.

\bibitem[P~Parambath et~al.(2016)P~Parambath, Usunier, and
  Grandvalet]{puthiya2016coverage}
Shameem~A P~Parambath, Nicolas Usunier, and Yves Grandvalet.
\newblock A coverage-based approach to recommendation diversity on similarity
  graph.
\newblock In \emph{RecSys}, pages 15--22. ACM, 2016.

\bibitem[Parambath et~al.(2018)Parambath, Vijayakumar, and
  Chawla]{puthiya2018saga}
Shameem~P Parambath, Nishant Vijayakumar, and Sanjay Chawla.
\newblock Saga: A submodular greedy algorithm for group recommendation.
\newblock In \emph{AAAI}, pages 3900--3908, 2018.

\bibitem[Santos et~al.(2010)Santos, Macdonald, and Ounis]{santos2010exploiting}
Rodrygo~LT Santos, Craig Macdonald, and Iadh Ounis.
\newblock Exploiting query reformulations for web search result
  diversification.
\newblock In \emph{Proceedings of the 19th international conference on World
  wide web}, pages 881--890. ACM, 2010.

\bibitem[Steck(2013)]{steck2013evaluation}
Harald Steck.
\newblock Evaluation of recommendations: rating-prediction and ranking.
\newblock In \emph{RecSys}, pages 213--220. ACM, 2013.

\bibitem[Su et~al.(2013)Su, Yin, Chen, and Yu]{su2013set}
Ruilong Su, Li'Ang Yin, Kailong Chen, and Yong Yu.
\newblock Set-oriented personalized ranking for diversified top-n
  recommendation.
\newblock In \emph{RecSys}, pages 415--418. ACM, 2013.

\bibitem[Vargas et~al.(2014)Vargas, Baltrunas, Karatzoglou, and
  Castells]{vargas2014coverage}
Sa{\'u}l Vargas, Linas Baltrunas, Alexandros Karatzoglou, and Pablo Castells.
\newblock Coverage, redundancy and size-awareness in genre diversity for
  recommender systems.
\newblock In \emph{RecSys}, pages 209--216. ACM, 2014.

\bibitem[Wasilewski and Hurley(2018)]{wasilewski2018}
Jacek Wasilewski and Neil Hurley.
\newblock Intent-aware item-based collaborative filtering for personalised
  diversification.
\newblock In \emph{UMAP}, pages 81--89. ACM, 2018.
\newblock ISBN 978-1-4503-5589-6.

\bibitem[Wu et~al.(2016)Wu, Liu, Chen, Yuan, Guo, and Xie]{Wu2016RMC}
Le~Wu, Qi~Liu, Enhong Chen, Nicholas~Jing Yuan, Guangming Guo, and Xing Xie.
\newblock Relevance meets coverage: A unified framework to generate diversified
  recommendations.
\newblock \emph{ACM Trans. Intell. Syst. Technol.}, 7\penalty0 (3), 2016.

\bibitem[Ziegler et~al.(2005)Ziegler, McNee, Konstan, and
  Lausen]{ziegler2005improving}
Cai-Nicolas Ziegler, Sean~M McNee, Joseph~A Konstan, and Georg Lausen.
\newblock Improving recommendation lists through topic diversification.
\newblock In \emph{WWW}, pages 22--32. ACM, 2005.

\end{thebibliography}
\end{document}